\documentclass[preprint]{revtex4}
\usepackage{amssymb}
\usepackage{makeidx}
\usepackage{amsmath}
\usepackage{graphicx}
\usepackage{dcolumn}
\usepackage{bm}

\begin{document}

\title{High harmonic generation in Landau-quantized graphene subjected to a
strong electromagnetic radiation }
\author{H.K. Avetissian}
\author{A.G. Ghazaryan}
\author{G.F. Mkrtchian}
\author{Kh.V. Sedrakian}
\affiliation{Centre of Strong Fields Physics, Yerevan State University, 1 A. Manukian,
Yerevan 0025, Armenia}

\begin{abstract}
We study nonlinear optical response of Landau quantized graphene to an
intense electromagnetic wave. In particular, we consider high harmonic
generation process. It is shown that one can achieve efficient generation of
high harmonics with strong radiation fields -- when the work of the wave
electric field on the magnetic length is larger than pump photon energy. At
that high harmonics generation process takes place for a wide range of the
pump wave frequencies and intensities even for significant broadening of
Landau levels because of impurities in graphene.
\end{abstract}

\maketitle



\section{INTRODUCTION}

Thanks to exotic nonlinear electromagnetic properties of graphene \cite{Nov1}%
, the latter is extensively considered as an active material for diverse
optical applications \cite{Nov2}. Note that the graphene is an effective
material for multiphoton interband excitation, wave mixing, and harmonic
generation processes \cite{25,26,27,28,30,Mer1,Mer2,Mer3,Mer4}. When a
static uniform magnetic field is applied perpendicular to the graphene
plane, the electron energy is quantized forming nonequidistant Landau levels
(LLs). As a consequence, in the graphene the anomalous quantum Hall effect
takes place \cite{Nov3,Zhang,GS,Goer}.

While most of the works have concentrated on static or linear optical
properties of the Landau-quantized graphene, one direction that has not been
fully explored is the nonlinear response of Landau-quantized graphene to a
strong coherent radiation. Linear optical response of graphene quantum Hall
system with the different aspects have been studied in Refs. \cite%
{MHA,AC1,AC2,AC3,AC4,LLL1,Ferreira}. Ultrafast carrier dynamics and carrier
multiplication in Landau-quantized graphene have been investigated in Refs. 
\cite{Carrier1,Carrier2}. In Refs. \cite{LLL1,LLL2} tunable graphene-based
laser on the Landau levels in the terahertz regime have been proposed.
Interesting effects also arise in the high frequency regime. As was shown in
Ref. \cite{MHA} the plateau structure in the quantum Hall effect in graphene
is retained, up to significant degree of disorder, even in the ac (THz)
regime, although the heights of the plateaus are no longer quantized. In
Refs. \cite{Mer5,Mer6} the nonlinear optical response of graphene and
semiconductor-hetero-structures to a moderately strong laser radiation in
the quantum Hall regime, in particular, radiation intensity at the 3rd
harmonic, as well as nonlinear Faraday effect have been investigated. It has
been shown that 3rd harmonic radiation intensity has a characteristic Hall
plateau structures that persist for a wide range of the pump wave
frequencies and intensities even for significant broadening of Landau levels
because of impurities in graphene. With further increase of pump wave
intensity and due to the peaks in the density of states one can also expect
enhancement of the high harmonics' radiation power in Landau-quantized
graphene. Hence, it is of interest to consider high harmonics generation
process in the ultrastrong wave-graphene coupling regime. Moreover, the
energy range of interest lies in the THz and Mid Infrared domain where
high-power generators and frequency multipliers are of special interest.

In the present work, a microscopic theory of the Landau-quantized graphene
interaction with strong coherent electromagnetic radiation is presented. Our
calculations show that one can achieve efficient generation of high
harmonics with strong radiation fields -- when the work of the wave electric
field on the magnetic length is Considerably larger than pump photon energy.
At that for optimization we investigate high harmonics generation process
depending on the pump wave frequency, intensity and broadening of LLs.

The paper is organized as follows. In Sec. II the Hamiltonian which governs
the quantum dynamics of considered process and the set of equations for a
single-particle density matrix are presented. In Sec. III, we numerically
solve obtained equations and consider high harmonics generation process.
Finally, conclusions are given in Sec. IV.

\section{BASIC MODEL}

The sketch of the considered scheme for harmonics generation is shown in
Fig. 1. The graphene sheet is taken in the $xy$ plane ($z=0$) and a uniform
static magnetic field is applied in the perpendicular direction. A plane
linearly polarized (along the $x$ axis) quasimonochromatic electromagnetic
radiation of carrier frequency $\omega _{0}$ and slowly varying envelope $%
E_{0}(t)$ interacts with the such system. Under these circumstances the
Hamiltonian of the system in the second quantization formalism in the
presence of a uniform time-dependent electric field 
\begin{equation}
E(t)=E_{0}(t)\cos \omega _{0}t  \label{e1}
\end{equation}%
can be presented in the form \cite{Mer5}:%
\begin{eqnarray}
\widehat{H} &=&\sum\limits_{n=-\infty }^{\infty
}\sum\limits_{m=0}^{N_{B}-1}\varepsilon _{n}\widehat{a}_{n,m}^{+}\widehat{a}%
_{n,m}  \notag \\
&&+\sum\limits_{n,n^{\prime }=-\infty }^{\infty
}\sum\limits_{m=0}^{N_{B}-1}E(t)\mathcal{D}_{n,n^{\prime }}\widehat{a}%
_{n,m}^{+}\widehat{a}_{n^{\prime },m},  \label{gr2}
\end{eqnarray}%
where $\widehat{a}_{n,m}^{\dagger }$\ and $\widehat{a}_{n,m}$ are,
respectively, the creation and annihilation operators for a carrier in a LL
state. The energy spectrum is given by 
\begin{equation}
\varepsilon _{n}=\mathrm{sgn}(n)\hbar \omega _{B}\sqrt{\left\vert
n\right\vert },  \label{energy}
\end{equation}%
where $\omega _{B}=\sqrt{2}\mathrm{v}_{F}/l_{B}$ plays the role of the
cyclotron frequency, $\mathrm{v}_{F}$ is the Fermi velocity. Here $n$ is the
LL index -- for an electron $n>0$ and for a hole $n<0$. The extra LL with $%
n=0$ is shared by both electrons and holes. The LLs are degenerate upon
second quantum number $m$ with the large degeneracy factor $N_{B}=\mathcal{S}%
/2\pi l_{B}^{2}$ which equals the number of flux quanta threading the 2D
surface $\mathcal{S}$ occupied by the electrons. The magnetic length is $%
l_{B}=\sqrt{c\hbar /eB}$ ($e$ is the elementary charge, $\hbar $ is Planck's
constant, $c$ is the light speed in vacuum, and $B$ is the magnetic field
strength). In Eq. (\ref{gr2}) $\mathcal{D}_{n,n^{\prime }}$ is the dipole
moment operator: 
\begin{equation}
\mathcal{D}_{n,n^{\prime }}=\frac{iel_{B}}{2\sqrt{2}}\left[ \varkappa
_{n,n^{\prime }}\mathbb{\delta }_{\left\vert n\right\vert ,\left\vert
n^{\prime }\right\vert +1}+\varkappa _{n^{\prime },n}\mathbb{\delta }%
_{\left\vert n\right\vert ,\left\vert n^{\prime }\right\vert -1}\right] 
\frac{\hbar \omega _{B}}{\varepsilon _{n^{\prime }}-\varepsilon _{n}},
\label{Dnn'}
\end{equation}%
where $\varkappa _{n,n^{\prime }}=\mathrm{sgn}(n)\sqrt{1+\delta _{n^{\prime
},0}}$. 
\begin{figure}[tbp]
\includegraphics[width=.52\textwidth]{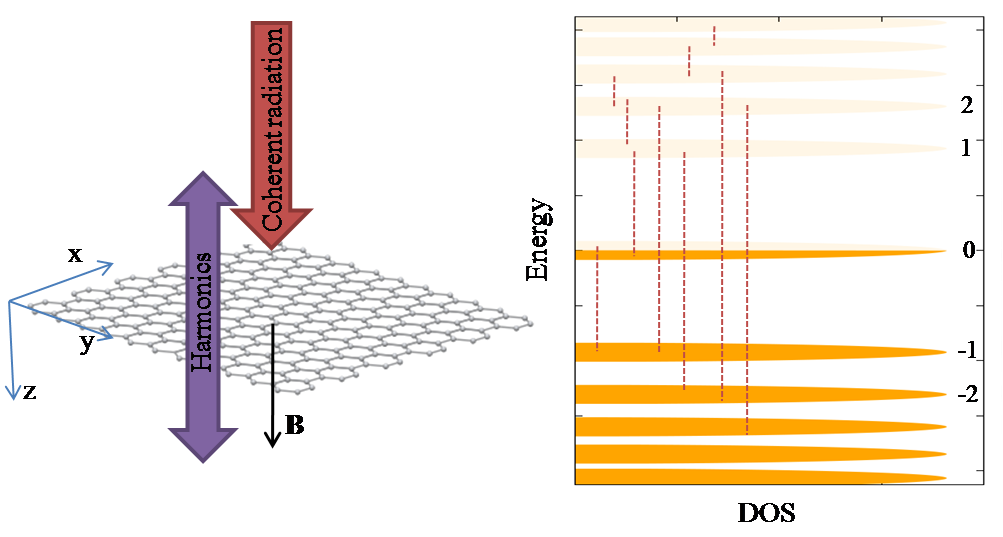}
\caption{ Schematic view of the considered scheme for generation of
harmonics. The graphene sheet is taken in the xy plane (z=0) and a uniform
static magnetic field is applied in the perpendicular direction. The optical
excitation via a linearly polarized coherent radiation pulse induces the
transitions between LLs which results surface currents and harmonics
generation. On the right hand side we show a typical density of states for
Landau-quantized graphene with some dipole allowed transitions.}
\end{figure}

Then we will pass to Heisenberg representation where operators obey the
evolution equation 
\begin{equation*}
i\hbar \frac{\partial \widehat{L}}{\partial t}=\left[ \widehat{L},\widehat{H}%
\right]
\end{equation*}%
and expectation values are determined by the initial density matrix $%
\widehat{D}$: $<\widehat{L}>=Sp\left( \widehat{D}\widehat{L}\right) $. In
order to develop microscopic theory of the nonlinear interaction of the
graphene QHE system with a strong radiation field, we need to solve the
Liouville-von Neumann equation for the single-particle density matrix%
\begin{equation}
\rho (n_{1},m_{1};n_{2},m_{2},t)=<\widehat{a}_{n_{2},m_{2}}^{+}(t)\widehat{a}%
_{n_{1},m_{1}}(t)>.  \label{grSPDM}
\end{equation}%
For the initial state of the graphene quasiparticles we assume an ideal
Fermi gas in equilibrium. According to the latter, the initial
single-particle density matrix will be diagonal, and we will have the
Fermi-Dirac distribution:%
\begin{equation}
\rho (n_{1},m_{1};n_{2},m_{2},0)=\rho _{F}\left( n_{1}\right) \frac{\delta
_{n_{1},n_{2}}\delta _{m_{1},m_{2}}}{1+\exp \left( \frac{\varepsilon
_{n_{1}}-\varepsilon _{F}}{T}\right) },  \label{grISPDM}
\end{equation}%
\begin{equation}
\rho _{F}\left( n_{1}\right) =\frac{1}{1+\exp \left( \frac{\varepsilon
_{n_{1}}-\varepsilon _{F}}{T}\right) }.  \label{FDD}
\end{equation}%
Including in Eq. (\ref{grISPDM}) quantity $\varepsilon _{F}$ is the Fermi
energy, $T$ is the temperature in energy units. As is seen from the
interaction term in the Hamiltonian (\ref{gr2}) quantum number $m$ is
conserved: $\rho (n_{1},m_{1};n_{2},m_{2},t)=\rho _{n_{1},n_{2}}\left(
t\right) \delta _{m_{1},m_{2}}$. To include the effect of the LLs broadening
we will assume that it is caused by the disorder described by randomly
placed scatterers. When the range of the random potential is larger than the
lattice constant in graphene, the scattering between $K$ and $K^{\prime }$
points in the Brillouin zone is suppressed and we can assume homogeneous
broadening of the LLs \cite{Ando}. The latter can be incorporated into
evolution equation for $\rho _{n_{1},n_{2}}\left( t\right) $ by the damping
term $-i\Gamma _{n_{1},n_{2}}\rho _{n_{1},n_{2}}\left( t\right) $ and from
Heisenberg equation one can obtain evolution equation for the reduced
single-particle density matrix:%
\begin{equation}
i\hbar \frac{\partial \rho _{n_{1},n_{2}}(t)}{\partial t}=\left[ \varepsilon
_{n_{1}}-\varepsilon _{n_{2}}-i\Gamma _{n_{1},n_{2}}\right] \rho
_{n_{1},n_{2}}(t)  \notag
\end{equation}%
\begin{equation}
-E(t)\sum\limits_{n}\left[ \mathcal{D}_{n,n_{2}}\rho _{n_{1},n}(t)-\mathcal{D%
}_{n_{1},n}\rho _{n,n_{2}}(t)\right] .  \label{grevol}
\end{equation}%
For the norm-conserving damping matrix we take $\Gamma _{n_{1},n_{2}}=\Gamma
\left( 1-\delta _{n_{1},n_{2}}\right) $, where $\Gamma $ measures the Landau
level broadening (see Fig. 1).

As is seen from Eqs. (\ref{grevol}) and (\ref{Dnn'}) in the Landau-quantized
graphene, wave-particle interaction can be characterized by the
dimensionless parameter $\chi _{0}=eE_{0}l_{B}/\hbar \omega _{0}$, which
represents the work of the wave electric field $E_{0}$ on the magnetic
length $l_{B}$ in units of photon energy $\hbar \omega _{0}$. Depending on
the value of this parameter $\chi _{0}$, one can distinguish three different
regimes in the wave-particle interaction process. Thus, $\chi _{0}<<1$
corresponds to the one-photon interaction regime, $\chi _{0}>>1$ corresponds
to the static field limit, and $\chi _{0}\succsim 1$ to the multiphoton
interaction regime. In this paper we consider just multiphoton interaction
regime ($\chi _{0}>1$) and look for features in the harmonic spectra of the
laser driven graphene.

\section{GENERATION OF HARMONICS AT THE MULTIPHOTON EXCITATION}

The optical excitation via a linearly polarized coherent radiation pulse
induces the transitions between LLs which results surface currents:%
\begin{equation*}
\mathcal{J}_{x}=-\frac{e\mathrm{v}_{F}}{\pi l_{B}^{2}}\sum\limits_{n,n^{%
\prime }}\rho _{n^{\prime },n}\left( \varkappa _{n,n^{\prime }}\mathbb{%
\delta }_{\left\vert n\right\vert ,\left\vert n^{\prime }\right\vert
+1}+\varkappa _{n^{\prime },n}\mathbb{\delta }_{\left\vert n\right\vert
,\left\vert n^{\prime }\right\vert -1}\right) ,
\end{equation*}%
\begin{equation}
\mathcal{J}_{y}=-\frac{ie\mathrm{v}_{F}}{\pi l_{B}^{2}}\sum\limits_{n,n^{%
\prime }}\rho _{n^{\prime },n}\left( \varkappa _{n^{\prime },n}\mathbb{%
\delta }_{\left\vert n\right\vert ,\left\vert n^{\prime }\right\vert
-1}-\varkappa _{n,n^{\prime }}\mathbb{\delta }_{\left\vert n\right\vert
,\left\vert n^{\prime }\right\vert +1}\right) .  \label{grcurry}
\end{equation}%
Here we have taken into account the spin and valley degeneracy factors $%
g_{s}=2$\ and $g_{v}=2$ and made summation over quantum number $m$ which
yields the degeneracy factor $N_{B}$. These currents have nonlinear
dependence on the pump wave field. At that one can expect intense radiation
of harmonics of the incoming wave-field in the result of the coherent
transitions between LLs. The harmonics will be described by the additional
generated fields $E_{x,y}^{(g)}$. We assume that the generated fields are
considerably smaller than the incoming field $\left\vert
E_{x,y}^{(g)}\right\vert <<\left\vert E\right\vert $. In this case we do not
need to solve self-consistent Maxwell's wave equation with Eq. (\ref{grevol}%
). To determine the electromagnetic field of harmonics we can solve
Maxwell's wave equation in the propagation direction with the given source
term:%
\begin{equation}
\frac{\partial ^{2}E_{x,y}^{(t)}}{\partial z^{2}}-\frac{1}{c^{2}}\frac{%
\partial ^{2}E_{x,y}^{(t)}}{\partial t^{2}}=\frac{4\pi }{c^{2}}\frac{%
\partial \mathcal{J}_{x,y}\left( t\right) }{\partial t}\delta \left(
z\right) .  \label{Max}
\end{equation}%
Here $\delta \left( z\right) $ is the Dirac delta function, $E_{x,y}^{(t)}$
is the total field. The solution to equation (\ref{Max}) reads%
\begin{equation*}
E_{x,y}^{(t)}\left( t,z\right) =E_{x,y}\left( t-z/c\right)
\end{equation*}%
\begin{equation}
-\frac{2\pi }{c}\left[ \theta \left( z\right) \mathcal{J}_{x,y}\left(
t-z/c\right) +\theta \left( -z\right) \mathcal{J}_{x,y}\left( t+z/c\right) %
\right] ,  \label{sol}
\end{equation}%
where $\theta \left( z\right) $ is the Heaviside step function with $\theta
\left( z\right) =1$ for $z\geq 0$ and zero elsewhere. The first term in Eq. (%
\ref{sol}) is the incoming wave. In the second line of Eq. (\ref{sol}), we
see that after the encounter with the graphene sheet two propagating waves
are generated. One traveling in the propagation direction of the incoming
pulse and one traveling in the opposite direction. The Heaviside functions
ensure that the generated light propagates from the source located at $z=0$.
We assume that the spectrum is measured at a fixed observation point in the
forward propagation direction. For the generated field at $z>0$ we have%
\begin{equation}
E_{x,y}^{(g)}\left( t-z/c\right) =-\frac{2\pi }{c}\mathcal{J}_{x,y}\left(
t-z/c\right) .  \label{solut}
\end{equation}%
Thus, solving Eq. (\ref{grevol}) with the initial condition (\ref{grISPDM})
and making summation in Eqs. (\ref{grcurry}) one can reveal nonlinear
response of the graphene. For the strong fields Eq. (\ref{grevol}) can not
be solved analytically and one should use numerical methods. For this
propose the time evolution of system (\ref{grevol}) is found with the help
of the standard fourth-order Runge-Kutta algorithm and for calculation of
Fourier transform of the functions $E_{x,y}^{(g)}\left( t\right) $ the fast
Fourier transform algorithm is used. For all calculations the temperature is
taken to be $T/\hbar \omega _{B}=0.02$ and Fermi energy is taken to be $%
\varepsilon _{F}=0$.

As is seen from Eqs. (\ref{grevol}) and (\ref{grcurry}), the spectrum
contains in general both even and odd harmonics. However, depending on the
initial conditions, in particular, for the equilibrium initial state (\ref%
{grISPDM}) and at the smooth turn-on-off of the wave field the terms
containing even harmonics cancel each other because of inversion symmetry of
the system and only the odd harmonics are generated. To avoid nonphysical
effects semi-infinite pulses with smooth turn-on, in particular, with
hyperbolic tangent 
\begin{equation}
E_{0}(t)=E_{0}\mathrm{tanh}(t/\tau _{r})  \label{envel}
\end{equation}
envelope is considered. Here the characteristic rise time $\tau _{r}$ is
chosen to be $\tau _{r}=20\pi /\omega _{0}$. Calculations show that for
harmonics $\left\vert E_{x}^{(g)}\right\vert >>\left\vert
E_{y}^{(g)}\right\vert $, that is harmonics are radiated with the same
polarization as incoming wave. Hence, we confine ourselves to only $%
E_{x}^{(g)}$. The emission strength of the $s$th harmonic will characterized
by the dimensionless parameter 
\begin{equation}
\chi _{s}=e\left\vert E_{x}^{(g)}\left( s\right) \right\vert l_{B}/\hbar
\omega _{0},  \label{khi}
\end{equation}%
where

\begin{equation}
E_{x}^{(g)}\left( s\right) =\frac{\omega _{0}}{2\pi }\int_{0}^{2\pi /\omega
_{0}}E_{x}^{(g)}\left( t\right) e^{is\omega _{0}t}dt.  \label{F1}
\end{equation}%
With the fast Fourier transform algorithm instead of discrete functions $%
\chi _{s}$ we calculate smooth function$\chi \left( \omega \right) $ and so $%
\chi _{s}=\chi \left( s\omega _{0}\right) $. Figures 2 and 3 show the
radiation spectrum via logarithm of the normalized field strength $\chi
\left( \omega \right) $ (in arbitrary units) versus pump wave intensity for
various frequencies. The LL broadening is taken to be $\Gamma =0.2\hbar
\omega _{B}$. From these figures we immediately notice maximums at the odd
harmonics and with the increase of the wave intensity the emission strengths
of the high harmonics become feasible, which are more favorable at low
frequencies.

\begin{figure}[tbp]
\includegraphics[width=.48\textwidth]{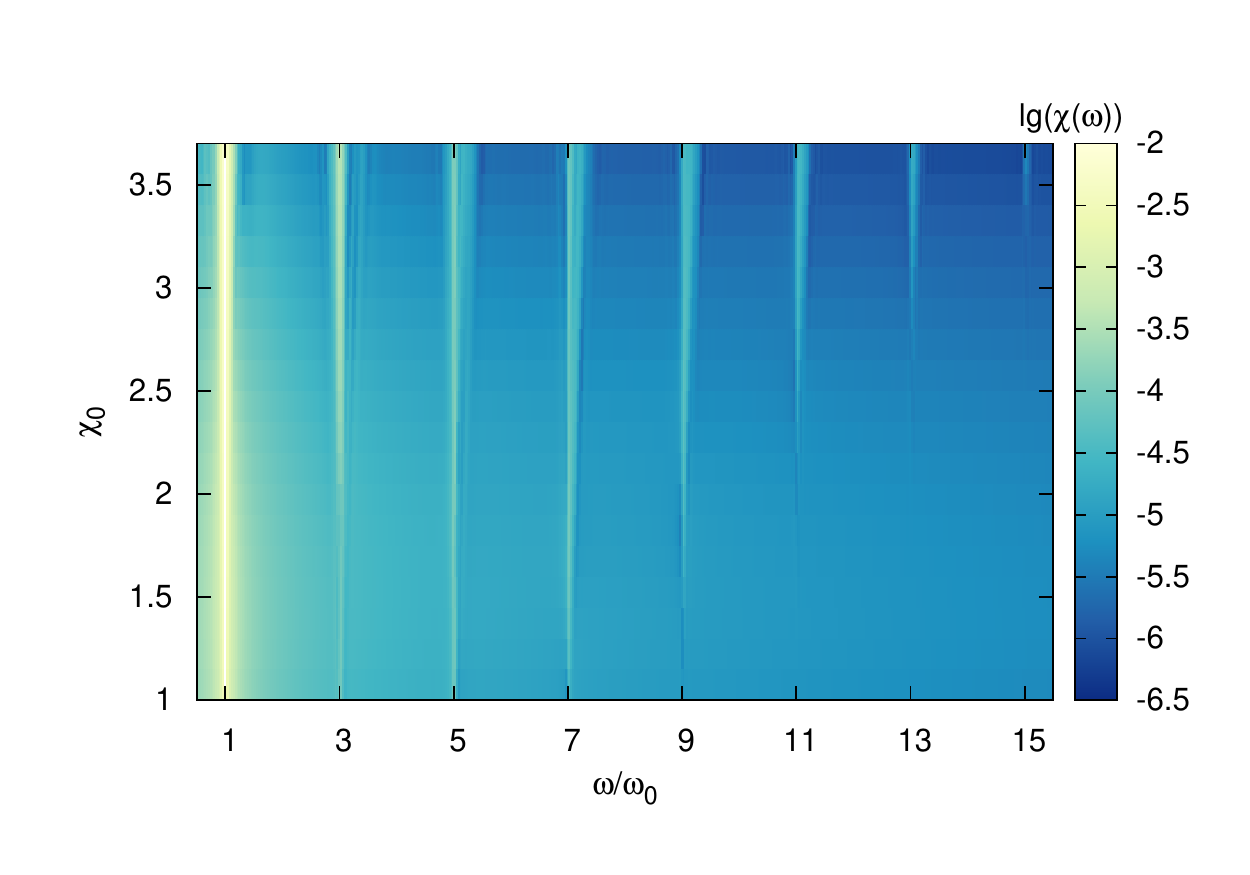}
\caption{The radiation spectrum via logarithm of the normalized field
strength $\protect\chi \left( \protect\omega \right) $ (in arbitrary units)
versus pump wave intensity with $\protect\omega _{B}/\protect\omega _{0}=1.5$%
. The LL broadening is taken to be $\Gamma =0.2\hbar \protect\omega _{B}$.}
\end{figure}
\begin{figure}[tbp]
\includegraphics[width=.48\textwidth]{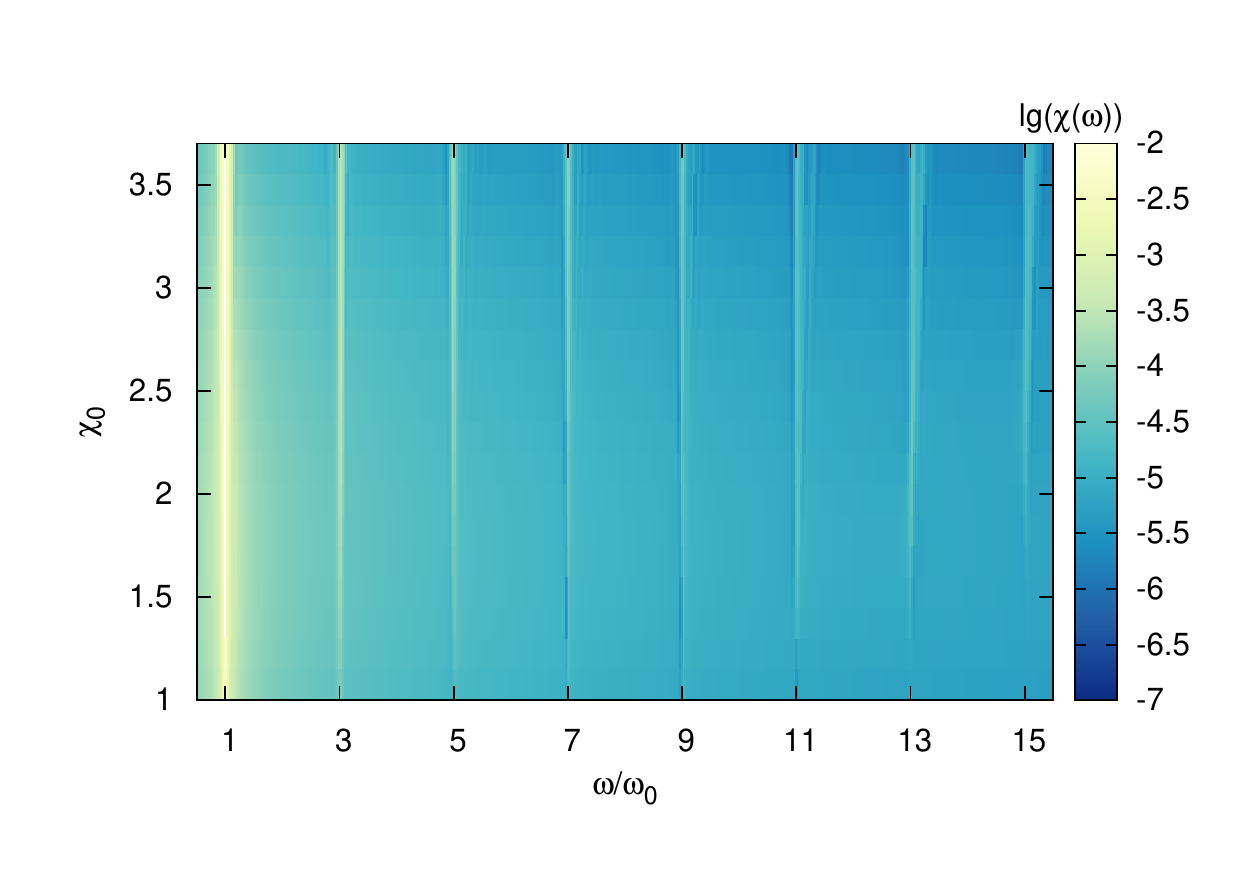}
\caption{Same as Fig. 2 but for $\protect\omega _{B}/\protect\omega _{0}=2$.}
\end{figure}

We further examine emission strengths of the 3rd, 5th, 7th, and 9th
harmonics for various pump wave frequencies and LL broadening at the fixed
value $\chi _{0}=4$, which is shown in Fig. 4. We can see from Fig. 4 that,
while the density of states broadens with a width $\sim \Gamma $ the
harmonics radiation rates are relatively robust and are feasible up to large 
$\Gamma $. Then, we see that emission of harmonics takes place for the wide
range of the pump wave frequencies. 
\begin{figure}[tbp]
\includegraphics[width=.52\textwidth]{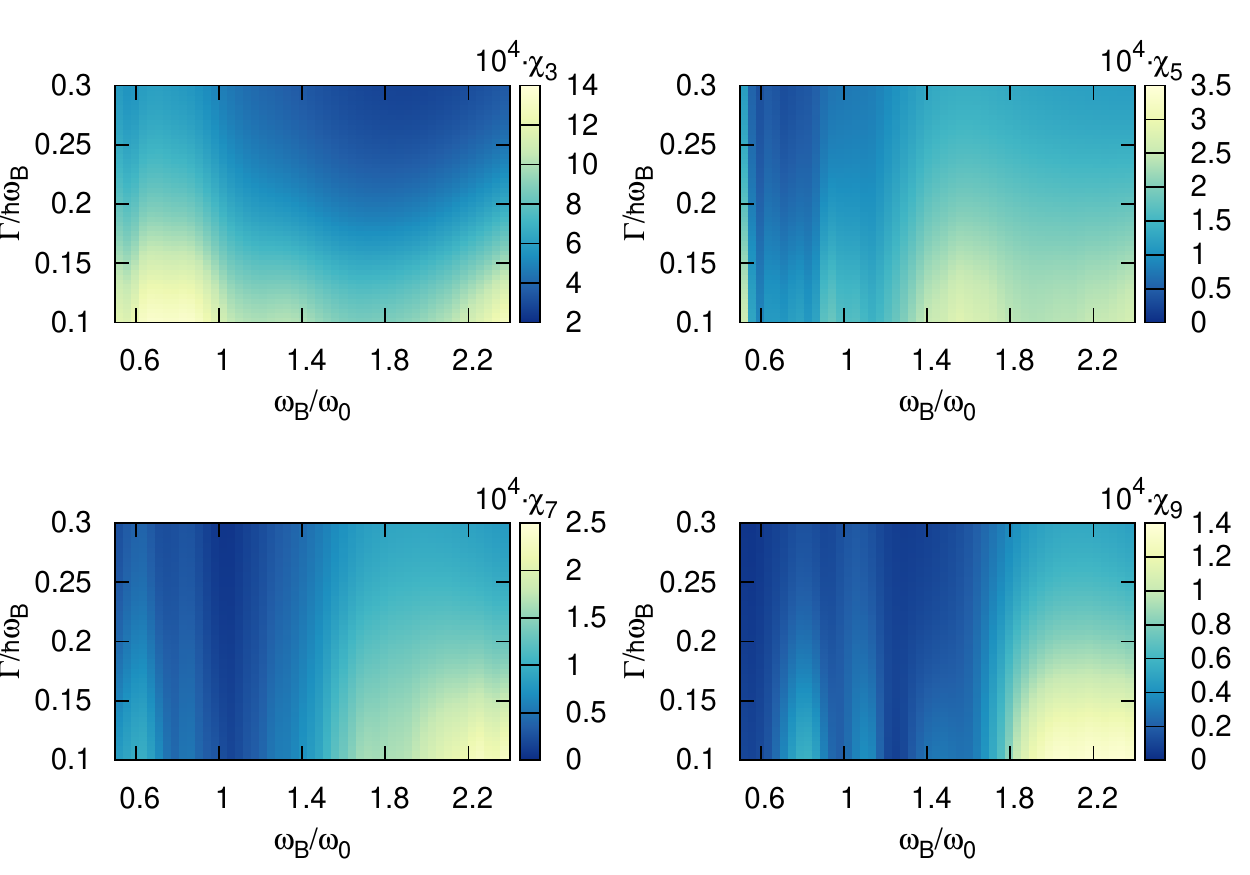}
\caption{Harmonic emission rate in Landau-quantized graphene at the
multiphoton excitation via $\protect\chi _{n}\equiv \protect\chi \left( n%
\protect\omega _{0}\right) $ for 3rd, 5th, 7th, and 9th harmonics versus LL
broadening and pump wave frequency (ratio $\protect\omega _{B}/\protect%
\omega $) for a wave of intensity $\protect\chi _{0}=4$.}
\end{figure}
\begin{figure}[tbp]
\includegraphics[width=.52\textwidth]{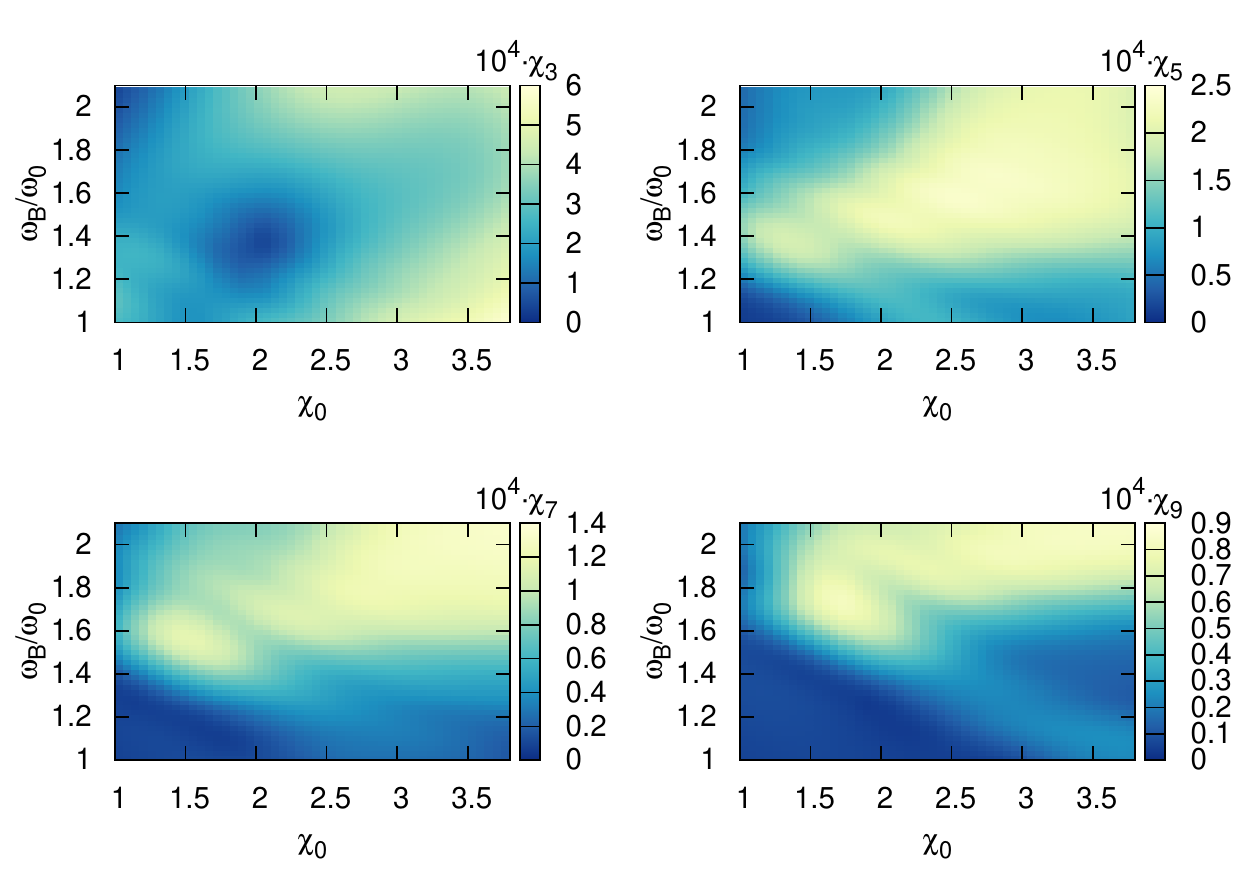}
\caption{Harmonic emission rate in Landau-quantized graphene at the
multiphoton excitation via $\protect\chi _{n}\equiv \protect\chi \left( n%
\protect\omega _{0}\right) $ for 3rd, 5th, 7th, and 9th harmonics versus
pump wave frequency (ratio $\protect\omega _{B}/\protect\omega $) and
intensity parameter $\protect\chi _{0}$. The LL broadening is taken to be $%
\Gamma =0.2\hbar \protect\omega _{B}$.}
\end{figure}
\ 

We also examine how the emission strengths of the 3rd, 5th, 7th, and 9th
harmonics behave depending on the pump wave intensity and frequency at the\
fixed value of LL broadening. The results of our calculations are shown in
Fig. (5). Thus, at $\chi _{0}\succsim 1$ we have intense radiation of
harmonics and optimal frequencies for pump wave are close to $\omega
_{B}/\omega _{0}\approx 2$.

\section{SUMMARY}

To summarize, we have presented a microscopic theory of the Landau-quantized
graphene interaction with coherent electromagnetic radiation towards high
harmonics generation. We have shown that the nonlinear optical response of
Landau-quantized graphene is quite large that persist for a wide range of
the pump wave frequencies and intensities $\chi _{0}>1$ even for significant
broadening of \ LLs because of impurities in graphene.

Let us consider the experimental feasibility of considered process. For the
pump wave field we will assume a $\mathrm{CO}_{2}$ laser with $\hbar \omega
\simeq 0.1\ \mathrm{eV}$. For the magnetic field we take $B=40\ \mathrm{T}$.
The average intensity of the wave for $\chi _{0}=3$ is $I_{0}\simeq 7\times
10^{8}$ $\mathrm{W/cm}^{2}$. It is clear that for the experimental
realization one needs multilayer epitaxial graphene \cite{epi}. We consider
experimentally achievable values $N_{L}\sim 50$ monolayers \cite{Layers}
with the film thickness $\sim 20$ $\mathrm{nm}$\textrm{.} Since film
thickness is much smaller than the considered wavelengths, the harmonics'
signal from all layers will sum up constructively. Thus, for the average
intensities of the harmonics we will have $I_{s}\simeq I_{0}N_{L}^{2}\chi
_{s}^{2}/\chi _{0}^{2}$. For the setup of Fig. 5 with the chosen parameters
the average intensities of the harmonics are estimated to be $%
I_{3}=7.2\times 10^{4}\ \mathrm{W/cm}^{2}$, $I_{5}=1.2\times 10^{4}\ \mathrm{%
W/cm}^{2}$, $I_{7}=4\times 10^{3}$ $\mathrm{W/cm}^{2}$, and $I_{9}\simeq
1.6\times 10^{3}$ $\mathrm{W/cm}^{2}$.

\begin{acknowledgments}
This work was supported by the RA MES State Committee of Science, in the
frames of the research project No. 15T-1C013.
\end{acknowledgments}

\end{document}